# Ambient Backscatter Communications in Mobile Networks: Crowd-Detectable Zero-Energy-Devices

Dinh-Thuy Phan-Huy[1], Dominique Barthel[1], Philippe Ratajczak[1], Romain Fara[1, 2], Marco di Renzo[2], Julien de Rosny[3]

[1]*Orange Labs Networks,* Châtillon, Meylan & Sophia-Antipolis, France
[2]*Université Paris-Saclay, CNRS, CentraleSupélec, Laboratoire des Signaux et Systèmes,* Gif-Sur-Yvette, France
[3]*ESPCI Paris, PSL University, CNRS, Institut Langevin*, Paris, France
{dinhthuy.phanhuy, dominique.barthel, philippe.ratajczak, romain.fara}@orange.com, marco.di-renzo@universite-paris-saclay.fr, julien.derosny@espci.fr

*Abstract*—In this paper, we introduce the new concept of Crowd-Detectable Zero-Energy-Devices. Such devices harvest solar or ambient light energy to power themselves, backscatter ambient waves to communicate, and are detected simultaneously by a smartphone connected to the network, and the network itself, as long as the device is close to the considered smartphone. Such device is a promising sustainable solution for the future of the Internet of Things. We describe an example of use case: asset tracking with zero added energy, zero new signals and zero new infrastructure in the network, thanks to the anonymous and transparent participation of smartphones connected to the wireless network. We then present our first prototypes of such devices that backscatter TV, 4G and 5G signals, and show our first experiments in challenging conditions and environments.

## I. Introduction

Each generation of mobile network, from the 2nd generation (2G) to the 5th generation (5G) has been improved in terms of spectral and energy efficiencies. Nevertheless, due to the fast traffic demand growth, the network energy consumption keeps growing [1]. Hence, there is strong need for finding sustainable solutions for wireless communications. Recently, ambient backscatter communications have been proposed [2]. In such communication systems, a tag sends a message to a reader without generating any radio-frequency (RF) wave. It backscatters ambient waves generated by an external source such as a television (TV) tower. In [3], such a system has been identified as a promising solution for Internet of Things (IoT) applications. In [4], real-time ambient backscatter communications have been demonstrated with a first tag prototype backscattering TV, 4th generation (4G) network waves and harvesting solar energy to power itself. However, the presented prototype was bulky and therefore very far from a product in terms of form factor. In [5], the experimental characterization of the performance of such link with the most basic detector (an energy-detector) has been performed, showing, performance fades in some localized areas. In [6-10], solutions to improve the performance have been proposed, by using massive multiple-input multiple-output (M-MIMO) antennas at the reader side [6] or at the source side [7], by using compact reconfigurable backscatters in polarization [8,9], or by assisting the communication with a reconfigurable intelligent surface in the context of a future 6th generation (6G) network [10].

However, up to now, a strongly attractive use case for mobile operators has not yet emerged. In this paper, for the first time, we introduce, in Section II, the concept of crowd-detectable zero-energy-device (CD ZED) based on ambient backscatters in mobile networks. We present the new and very promising use-case of "asset tracking out-of-thin-air" in Section III. We also present our first prototypes in Section IV. Finally, in Section V, we describe some first experimentations and demonstrations presented at the 2021 Orange Labs Research Exhibition [11] and the 2021 Mobile World Congress [12].

## II. CD ZED Concept

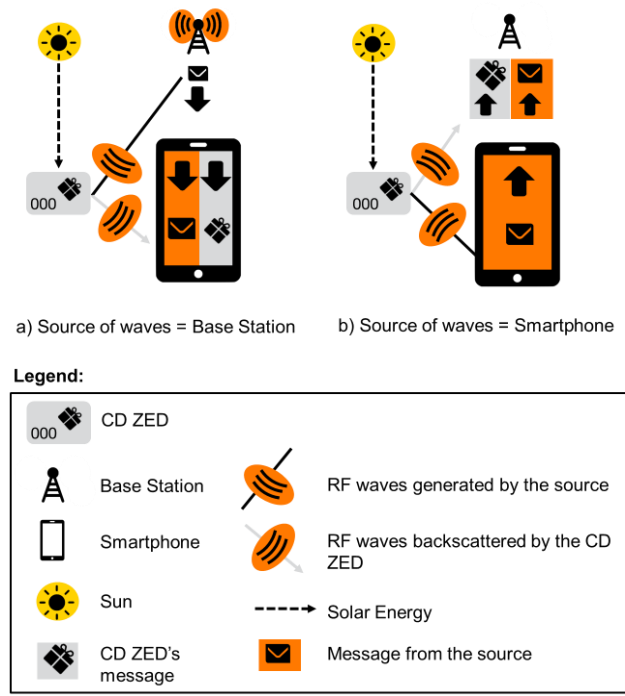

Fig. 1. CD ZED concept

A CD ZED harvests solar or ambient light energy to power itself, backscatter ambient waves to communicate, and is detected simultaneously by a smartphone connected to the network, and the network itself, as long as the device is close to the considered smartphone. As illustrated in Fig. 1, each time a CD ZED gets close to a smartphone connected to the network, it backscatters the waves that are generated by the network and that are intended to the smartphone. Simultaneously, it backscatters the waves that are generated by the smartphone and that are intended to the network. The smartphone can simultaneously demodulate the data coming from the network and act as a RF tag reader, by detecting the tag message through an analysis of the propagation channel variations. Similarly, the network can simultaneously demodulate the signal coming from the network, plus act as a RF tag reader, by detecting the tag's message in the channel variations. Hence, the tag's detection can be associated to the smartphone location. In other terms, when a tag gets close to a smartphone, it is automatically located and time-stamped by the smartphone and the network.

By organizing an anonymous participation of all its smartphone customers, an operator can therefore track the locations of any tags that "bump" into its float or "crowd" of smartphones. To protect its smartphone customers' personal data, it is important that the mobile operator guarantees collects the acceptance from each customer of a participation agreement (through an electronic form for instance). This agreement must include guarantees from the operator, that the customer's participation will remain completely anonymous. More precisely, once a tag is detected, located and time-stamped, no personal data linked to the smartphone nor the customer identity must be stored.

Such device is a promising sustainable solution for the future of the IoT. Indeed, it enables a mobile network operator to develop IoT use cases without spending more energy, without emitting new waves and without deploying additional infrastructure in the network.

### III. ASSET-TRACKING OUT-OF-THIN-AIR

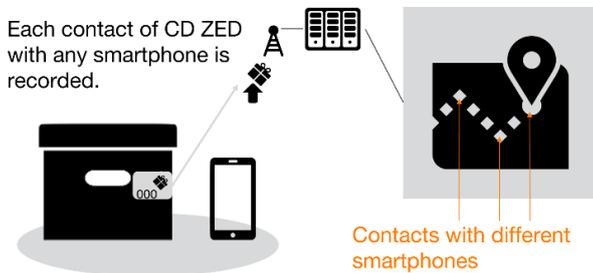

Fig. 2. Asset tracking with zero added energy, zero new signals and zero new equipment in the network

In this section, we describe an example of promising use case: "Asset-Tracking Out-Of-Thin-Air", i.e. asset tracking with zero added energy, zero new signals, and zero new infrastructure in the network. As illustrated in Fig. 2, a transportation company puts a CD ZED on a package. Each time the package gets close to a smartphone, it is automatically located, time-stamped by the network. Thanks to the anonymous and transparent participation of smartphones connected to the wireless network (described in Section III), the trajectory of the package can be tracked in outdoor environments and every locations where no Radio Frequency Identification (RFID) portals or readers are deployed.

### IV. ZERO-ENERGY-DEVICES PROTOTYPES

In this section, we present our first prototypes of CD ZEDs: compact solar tags with a fixed message.

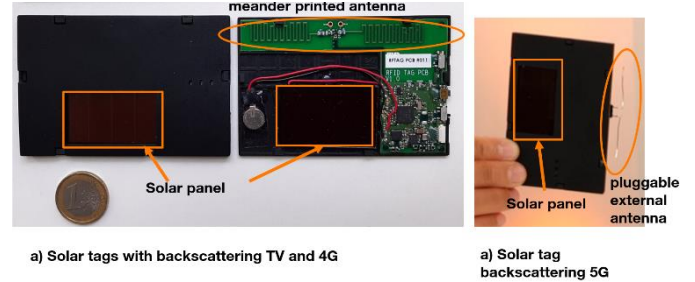

Fig. 3. Prototypes of CD ZEDs: business card size solar tags recycling TV/4G (between 480 and 800 MHz) and 5G signals (at around 3.5 GHz)

Our tag is the size of a business card and is mainly composed of a dipole antenna connected to a RF switch, itself being controlled by a MSP430. A BQ25570 battery charger is used to harvest micro watt energy from an indoor solar panel AM-1454CA and to store it in a 3V coin cell battery. Hence, a tag self-powers with solar or ambient indoor light. A tag repeatedly sends a black and white pixel image of 8 lines and 11 columns (88 pixels in total), each black or white pixel being coded by a bit. A synchronization sequence of 8 bits is appended to the 88 bit-sequence corresponding to the image. The whole 96 bits sequence is sent using FM0 modulation, hence with 96×2=192 FM0 bits. The RF switch connects the two branches of the dipole when the FM0 bit is equal to zero, and disconnects them otherwise. Depending on the impedance that loads the dipole antenna, the tag backscatters a different level of ambient signal. To guarantee a good contrast between the two states, we have chosen the Infineon BGS12WN6 wideband SPDT Diversity Switch that provides high isolation and low insertion loss at frequencies below 6 GHz. Finally, the tag's rate is computed as follows. The tag switches every Ts seconds, sends Fs=1/Ts FM0 symbols per second and Fb=1/(2Ts) bits (or pixels) per second. The setting of Ts depends on the ambient source. For instance, in 5G, Ts must be set to a larger value that the Time Division Duplex frame duration T_5G of 1 ms, to avoid any confusion between the tag's symbol period and the frame period. Ts setting for each type of ambient source is given in Table 1.

Regarding the dipole antennas two options are available. Either a printed meander antenna as illustrated in Fig. 3-a) or an external pluggable dipole antenna is used as illustrated in Fig. 3-b). Fig. 4-a) and Fig. 4-b) illustrate the external dipole antenna based tag and meander based tag solutions, respectively. Fig. 4 also includes plots of the corresponding received signals at the reader side, when the tag is in short-circuit or open circuit state, respectively, using an High Frequency Simulation Software (HFSS) tool. In these simulations, the source and the reader are dipole antennas located above the tag, in the boresight of the antenna of the tag, with the same polarization than the tag, to maximize the tag's detection. As illustrated in Fig. 4), the advantage of the dipole is that the contrast between the two states is better and the bandwidth in which the contrast is significant is larger. The drawback of this solution, is the size of the tag (it reaches around 20 cm at 700 MHz). The meander based antenna tag is advantageously more compact. However, the contrast of the meander antenna is weaker and the bandwidth in which the contrast is significant is much narrower. Finally, the central frequency of the meander dipole antenna based solution is shifted in frequency, compared to the external dipole solution. In this case, an inductance is inserted in the antenna matching network to configure the central frequency of the tag.

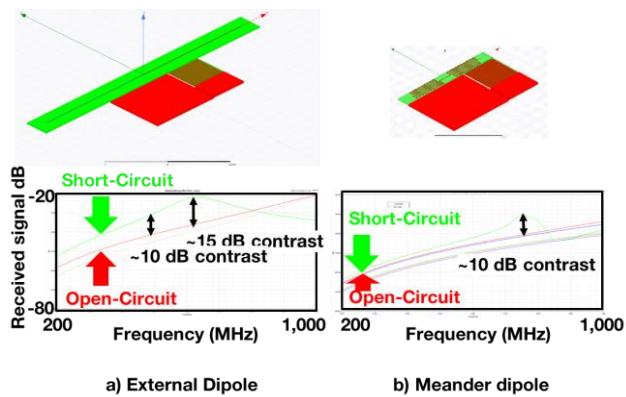

Fig. 4. Full-wave simulation using HFSS of (a) external pluggable dipole antenna based tag and (b) meander antenna based tag.

## V. EXPERIMENTS AND DEMONSTRATIONS

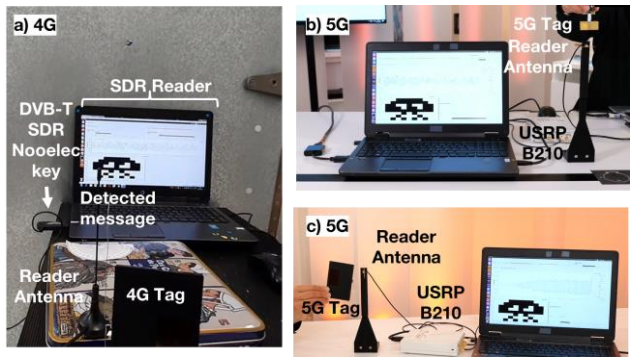

Fig. 5. Ambient backscatter experiment with a) 4G business card size solar tag (**video** [13]), b) 5G tag at 3.65 GHz and Ericsson small cell from Orange 5G commercial network as source GHz (**video** [14]); c) 5G business card size solar tag at 3.65 GHz (**video** [15]).

In this section, we present our first experiments with the compact solar tags presented in Section IV. In these experiments, the tag backscatters ambient signals from TV towers and 4G or 5G base stations from Orange commercial mobile networks. Our prototype of smartphone (or RF reader) is based on a Software Defined Radio (SDR) GNU platform running on a portable computer (PC), and initially developed for TV and 4G only, in [4]. Some adaptations presented hereafter, have been introduced to support 5G.

Two slightly different hardware setups have been designed for low bands (4G around 700 MHz and TV around 500 to 700 MHz) and for high bands (5G at 3.65 GHz). More precisely, as illustrated in Fig. 5-a), our reader for 4G and TV, is connected to an RTL Terrestrial Digital Video Broadcast (DVB-T) SDR Nooelec key connected to a monopole antenna for below 2 GHz bands. As illustrated in Fig. 5-b) and c), our reader for 5G is connected to a Universal Software Radio Peripheral (USRP) B210 board connected to a dipole antenna for 3.65 GHz.

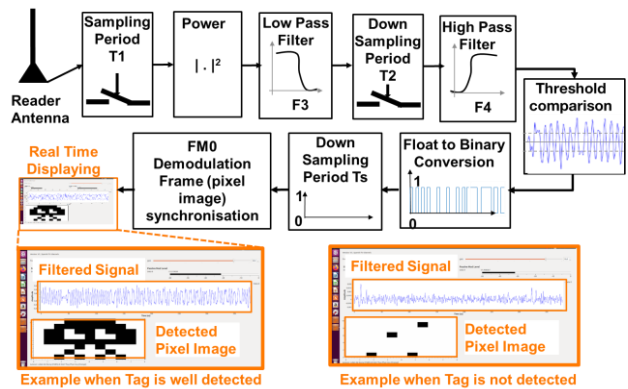

Fig. 6. Reader's algorithm with configurable parameters

The reader demodulates in real-time the tag's message thanks to the most simple signal processing method: the energy-detector or threshold detector used in [2]. The detector can be configured for various ambient sources. More precisely, as illustrated in Fig. 6, this detector samples the received signal with period T1. Then, complex samples are converted into power samples, filtered with a low pass filter with F3 cut-off frequency, down-sampled with period T2, filtered with a high pass filter with F4 upper cut-off frequency, compared to a threshold (the moving average over a time-window Ta), and converted into binaries ('1' if the sample exceeds the threshold and '0' otherwise). The resulting binary sequence is down-sampled with the tag's symbol Ts and demodulated assuming FM0. Finally, the synchronization sequence is detected (by correlation) and the consecutive 88 bit sequence is displayed as an 8 by 11 pixels image. The setting of the reader parameters T1 (F1=1/T1), T2 (F2=1/T2), F3, F4 and Ta are optimized depending on the ambient source, according to Table 1. For 5G, as the traffic is very bursty, Ta/Ts is not chosen as large as for TV and 4G. Note that such detector is non-coherent, without a priori knowledge on the ambient source, and is low complexity enough to be run in real-time on our PC, as illustrated in filmed experiments [13][14][15].

TABLE I. PARAMETERS VALUES

| Parameter | TV/4G | 5G |
|---|---|---|

| Ts, Fs | 2.7 ms, 370 Hz | 10.8 ms (>> T_5G), 93 Hz |
| --- | --- | --- |
| 2Ts, Fb | 5.4 ms, 185 Hz | 21.6 ms, 92 Hz |
| F3 (>Fs), F4 (<Fb) | 500 Hz, 50 Hz | 100 Hz, 1 Hz |
| T1, F1 (>Fs) | 1 µs, 1 MHz | |
| T2, F2 (>Fs) | 0.5 ms, 2 kHz | |
| Ta (>> Ts) | 50 ms | |

Fig. 7-a) and b) illustrate indoor experiments with ambient TV and business card size solar tags using a pluggable external dipole antenna (optimized for 540 MHz), at different floors of office buildings. In these experiments, around 4 meters tag-to-reader range was reached with the simplest detector, even though the signal is attenuated by heat-blocking window glass. This is due to the fact that the tag's rate Fb is extremely slow. Also, the experiment was reproduced successfully with the same antenna for various TV channels with carrier frequencies lying between 482 MHz and 700 MHz, confirming that the pluggable dipole antenna tag has a large bandwidth. Finally, Fig. 7-c) illustrates an ambient backscatter communication experiment with ambient TV in challenging conditions: in mobility, underground in a metro-train.

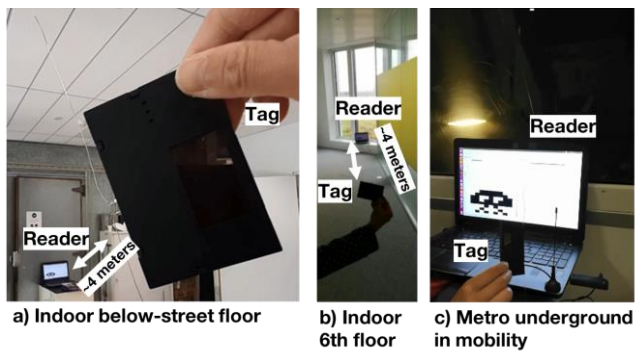

Fig. 7. Ambient TV backscattering experiments in indoor at a) street floor (**video** [16]), and b) 6[th] floor of an office building (**video** [17]), also in mobility c), underground, in metro train line 13, in a tunnel between Porte de Vanves and Malakoff Plateau de Vanves Stations (**video** [18]).

## VI. CONCLUSION

In this paper, we have introduced for the first time the concept of crowd-detectable zero-energy-devices, and the use case of asset-tracking out-of-thin-air. We have also presented first prototypes of such devices – business card size solar tags – and tested them with various ambient sources (signals from TV, 4G and 5G ambient commercial networks), and in various conditions, such as indoor and mobility underground. These experiments show that ambient backscattering is a promising technique for mobile networks. Next studies will focus on field trials with improved signal processing at the reader side, and also, extension to future 6[th] generation networks.

## VII. ACKNOWLEDGMENT

We thank our colleagues M.-A. Mouilleron, P. Legay and G. Tardiveau for their expertises and contribution on use cases identifications. We thank the Orange teams of the Orange Labs Research Exhibition and the Mobile World Congress.